\documentclass{optica-article}

\journal{opticajournal}

\begin{document}
\title{Sub-0.1 degree phase locking of a single-photon interferometer}

\author{Vojtěch Švarc,\authormark{1,*} Martina Nováková,\authormark{1} Michal Dudka,\authormark{1} and Miroslav Ježek\authormark{1}}

\address{\authormark{1}Department of Optics, Faculty of Science, Palack\'y University, 17.\ listopadu 12, 77146 Olomouc, Czech Republic}
          
\email{\authormark{*}svarc@optics.upol.cz} 



\begin{abstract}
We report a single-photon Mach-Zehnder interferometer stabilized to a phase precision of 0.05 degrees over 15 hours. To lock the phase, we employ an auxiliary reference light at a different wavelength than the quantum signal. The developed phase locking operates continuously, with negligible crosstalk, and for an arbitrary phase of the quantum signal. Moreover, its performance is independent of intensity fluctuations of the reference. Since the presented method can be used in a vast majority of quantum interferometric networks it can significantly improve phase-sensitive applications in quantum communication and quantum metrology.
\end{abstract}

\section{Introduction}
Interferometers locked to a specific phase are widely used in many fields of photonics research and technology, such as quantum information processing \cite{ma2011,mikova2012}, quantum metrology \cite{Giovannetti2011,LIGO2013squeezing,Thekkadath2020}, quantum communication \cite{Xavier2011,Xavier2015,Toliver2015,Cho2016,Xu2019}, and tests of fundamental physics \cite{ma2012,Vallone2018,Svarc2020}. In these applications, precise phase control of single-photon interferometers and photonic routers is crucial \cite{Toliver2015,Vallone2018,Svarc2019}. For instance, the phase stability determines the amount of quadrature squeezing required for protocols in quantum metrology \cite{Takeno2007,Eberle2013}. In quantum cryptography and quantum communication, phase instabilities increase the bit error rate, which deteriorates the speed and security \cite{Makarov2004}. Generally, sub-degree stability is considered to be sufficient for most applications except for ultra-sensitive ones such as loop-based protocols where the phase error cumulates each roundtrip \cite{Furusawa2017,takeda2019,Svarc2019}.

Although in classical interferometry, the sub-degree phase stabilization is feasible, at the single-photon level it is extremely challenging due to Poissonian photodetection noise. As a consequence, the typical precision of single-photon phase locking is a few degrees, and the response is slower than 0.1~s \cite{Makarov2004,Huntington2005}. An efficient strategy to overcome the photodetection noise is to lock the phase via a bright reference light co-propagating with the signal photons. To address the signal and the reference individually, they must differ in a selected parameter. The universal approach, suitable for both free-space and fiber setups, is to differentiate the signal and the reference by wavelength \cite{Xavier2011,Toliver2015,Cho2016}. An alternative approach, suitable for free-space setups only, is to add a small transversal displacement between the signal and the reference beams \cite{ma2011,ma2012}. 

Optimal phase locking should possess the following features: excellent phase stability, long-term performance, continuous operation, and minimal crosstalk between the signal and the reference. Most of the works in the field of single-photon interferometer phase locking did not satisfy more than two of these requirements \cite{Makarov2004,Huntington2005,Suzuki2006,Takeno2007,Cho2009,
Xavier2011}. Specifically, most of them are limited to short-term operation and few-degree stability. Two works demonstrated continuous operation, reasonably low crosstalk, and 0.2-degree stability in the short term \cite{Toliver2015} or a 4-degree stability in the long term \cite{yanikgonul2019}. However, achieving all of the desired features simultaneously has not been reported so far.

Here we demonstrate continuous sub-0.1 degree phase locking of a single-photon Mach-Zehnder interferometer (MZI) in the dual-wavelength configuration. To lock the phase, we use spectrally stable 1.5~nW reference light adaptively corrected for intensity fluctuations, and as a result, we reach exceptional phase stability of 0.05 degrees for 15~hours demonstrated for the signal. Moreover, using integrated electro-optic modulators embedded in MZI, we achieve sub-ns phase switching. The single-photon signal is virtually free of crosstalk from the reference and exhibits visibility of 99.6$\%$ for 3~nm spectral width, enabling short-pulse operation. These properties make our approach directly applicable in many quantum protocols requiring fast photonic routing or low-latency feedforward operation \cite{Svarc2019,Svarc2020}. Furthermore, the presented phase stabilization can facilitate demanding loop-based protocols \cite{Rohde2014,Pan2017,Furusawa2017,Svarc2019,takeda2019} and enable ultra-stable operation of time-bin quantum communication protocols \cite{Vallone2018,Xavier2011,Xavier2015,Toliver2015}.

\section{Methods}
The phase lock is commonly carried out by an active feedback loop consisting of a detector, a proportional-integral-derivative (PID) controller, and a phase modulator. The principle is following: A phase fluctuation is imprinted to the detected light as an intensity fluctuation. Subsequently, the produced electronic signal is evaluated by the PID controller, and an appropriate voltage is fed into the phase modulator in order to compensate for the initial phase fluctuation. The precision and speed of this approach depend on noise, frequency response, and dynamic range of the feedback loop. However, there are additional challenges one has to count on. In this section, we will focus on intensity and wavelength-related issues and ways how to suppress them. Although these issues are not commonly addressed, they can significantly limit the performance of the active feedback loop, especially while aiming at long-term sub-degree stability.

\begin{figure}[h!]
\centering\includegraphics[width=0.9\linewidth]{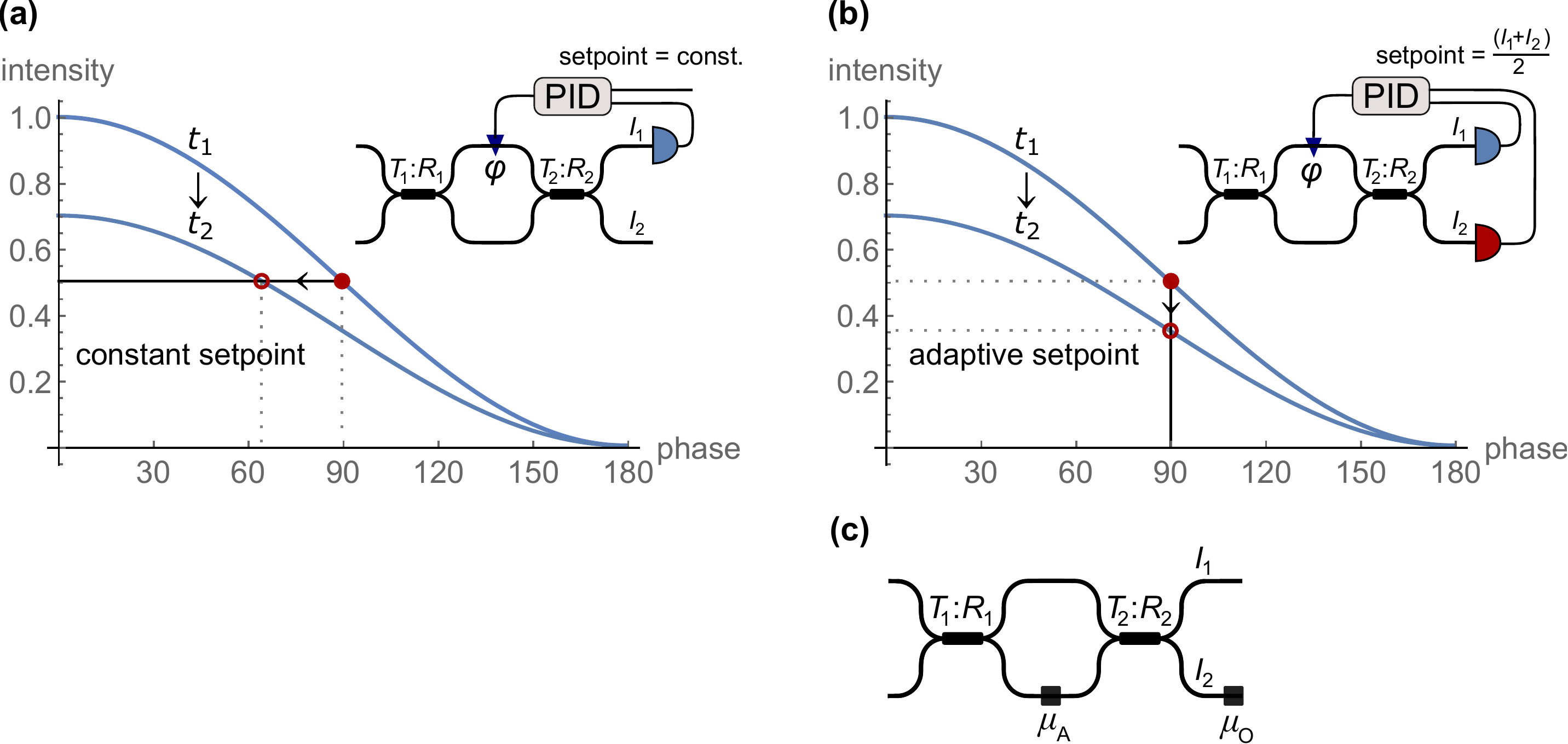}
\caption{Phase locking with (a) constant setpoint and (b) adaptive setpoint. Phase versus intensity (interference fringe) is plotted for time $t_1$ and $t_2$. The phase is set to 90 degrees in $t_1$, and then in $t_2$, the total intensity decreases. (a) If the phase is locked using one detector and a constant setpoint, the phase lock cannot distinguish between intensity and phase drift. Consequently, decreased intensity in $t_2$ leads to a phase locking error. (b) By detecting both outputs, the setpoint can be adaptively corrected for the intensity fluctuations. Therefore, the phase is kept constant despite decreased intensity in $t_2$. (c) An extended model of MZI including LSD parameters $\mu_\mathrm{O}$ and $\mu_\mathrm{A}$.}
\label{fig:ad_set}
\end{figure}

Conventionally, the PID controller evaluates the actual phase error by comparing the intensity detected at a single interferometer output with a constant setpoint. However, as depicted in Fig.~\ref{fig:ad_set}(a), this solution cannot distinguish between intensity and phase fluctuations. Therefore the setpoint does not correspond to the desired phase if the intensity of the reference fluctuates. This issue can be overcome by evaluating the setpoint using the actual intensity of the reference. As shown in Fig.~\ref{fig:ad_set}(b), this is possible by monitoring both interferometer outputs, summing their outcomes, and using a fraction of the sum as the setpoint. Apart from suppressing the influence of intensity fluctuations, this adaptive setpoint improves the signal-to-noise ratio (SNR) by a factor of 2 since two independent noisy detectors are averaged.

Although the adaptive setpoint eliminates efficiently the total intensity changes, it does not sufficiently respond to local intensity fluctuations caused by an asymmetric misalignment of the setup (e.g. detuning in a single arm). This effect, which we call local setup detuning (LSD), is typically much weaker than the total intensity drift, but it is harder to determine and correct for. LSD spoils the phase estimation, which consequently leads to a phase error since the setpoint no longer corresponds to the desired phase. We divide LSD into two types described by parameters $\mu_\mathrm{O}$ and $\mu_\mathrm{A}$ as illustrated in Fig.~\ref{fig:ad_set}(c).

The first type of LSD is relative detuning between interferometer outputs which is equivalent to a time-dependent loss $\mu_\mathrm{O}$ at a single output port. The phase estimation $\varphi_\mathrm{EST}$ using output intensities $I_1$ and $I_2$ is described as

\begin{equation}
\varphi_\mathrm{EST}=\arccos \left[ \dfrac{ \dfrac{I_1}{V_2} - \dfrac{I_2}{V_1}  \left(1-\mu_\mathrm{O} \right)}{I_1+I_2\left(1-\mu_\mathrm{O} \right)  } \right],
\end{equation} 
where $V_1$ and $V_2$ are visibilities at the corresponding output ports. If $\mu_\mathrm{O}=0$, $\varphi_\mathrm{EST}$ is equivalent to the real phase $\varphi$. However, if $\mu_\mathrm{O}\neq0$, the phase is estimated with an error given as $\varphi_\mathrm{EST}-\varphi$. For example, let us assume that $\mu_\mathrm{O}=1\%$ is introduced in the reference. Then the incorrect phase estimation causes a phase error of 0.3 degrees. Here we assume a high-visibility regime and phase locked close to $\frac{\pi}{2}$ (parameters relevant to our setup). This example illustrates that even small detuning would cause non-negligible phase error.

The second type of LSD is described by detuning of a single arm in the interferometer, equivalent to time-dependent loss $\mu_\mathrm{A}$, which impacts the visibility and intensity distribution between the outputs. As a consequence, the phase estimation is biased since we have no access to the actual visibility during the stabilization process. To describe this effect, it is necessary to express the intensities and visibilities in Eq.~(1) using transmittances $T_1$ and $T_2$ of beam splitters employed. For simplicity, let us assume perfect coherence and $\mu_\mathrm{O}=0$. Then the corresponding equation for the estimated phase in MZI reads
\begin{equation}
\varphi_\mathrm{EST}=\arccos\left[ \frac{T_2 (1-T_2)(1-2T_1)\mu_\mathrm{A}+2 \sqrt{T_1 T_2 (1-T_1)(1-T_2)(1-\mu_\mathrm{A})} \cos \varphi}
{2 \sqrt{T_1 T_2 (1-T_1) (1-T_2)}(1-T_1 \mu_\mathrm{A})}    \right].
\end{equation}
From this analysis, the phase estimation error is negligible if $\mu_\mathrm{A}$ is small (less then $10\%$) and the first beam splitter is balanced perfectly ($T_1=0.5$). Unfortunately, fiber couplers show a strong dependence on wavelength, so the splitting ratio for either the signal or the reference typically differs from 50:50. We will demonstrate this effect using parameters relevant to our setup. Let us assume that the reference phase is locked at $\frac{\pi}{2}$ and the interferometer arm losses are relatively detuned by $1\%$, which corresponds to $\mu_\mathrm{A}=1\%$. This detuning can potentially impact both the signal and the reference. But in our case, the fiber couplers are designed for the signal with the wavelength 810~nm, thus $T_{1,2}\approx0.5$ and the signal phase is estimated correctly. However, for the reference with the wavelength 840~nm, the fiber couplers are unbalanced ($T_{1,2}=0.35$). According to Eq.~(2), the reference phase is estimated with a 0.1-degree error. Consequently, the phase is locked with a 0.1-degree error.
 
So far, we have discussed phase errors caused by intensity fluctuations. Now let us briefly discuss the impact of wavelength fluctuations. As the phase $\varphi$ and the wavelength $\lambda$ are fundamentally connected by equation $\varphi=2\pi \frac{\Lambda}{\lambda}$, $\Lambda$ being optical path difference, wavelength instabilities in general affect the phase. In the dual-wavelength configuration, the phase error is present if the phase between the signal and the reference drifts relatively. The relative phase drift $\varphi_\mathrm{s}-\varphi_\mathrm{r}$ for small wavelength drift $\Delta\lambda$ is described as

\begin{equation}
\varphi_\mathrm{s}-\varphi_\mathrm{r}=2\pi\left(\frac{\Lambda_\mathrm{s}}{\lambda_\mathrm{s}^{2}}\Delta\lambda_\mathrm{s}-\frac{\Lambda_\mathrm{r}}{\lambda_\mathrm{r}^{2}}\Delta\lambda_\mathrm{r} \right),
\end{equation}
where indices s and r denote the signal and the reference, respectively. To avoid the relative phase drift, the wavelengths have to be locked, or the interferometer needs to be balanced perfectly. The latter also requires precise dispersion compensation since the zero optical path difference is necessary for both drifting wavelengths.

\section{Experiment}
\begin{figure}[h!]
\centering\includegraphics[width=\linewidth]{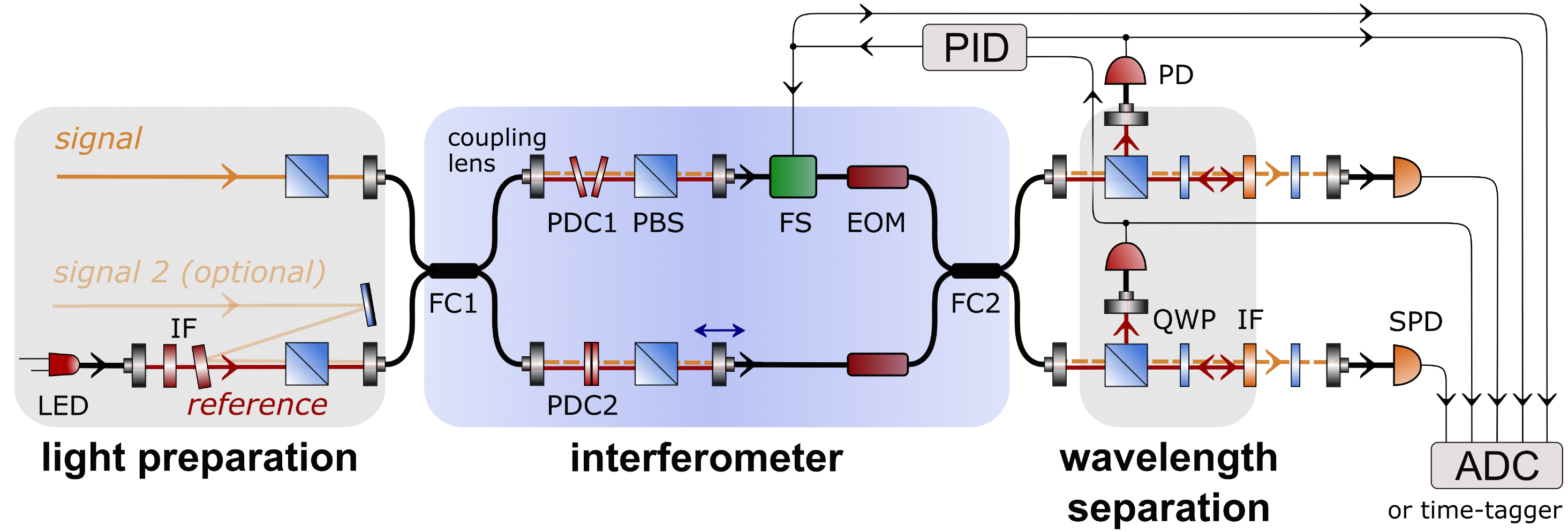}
\caption{Experimental setup for an interferometer with active phase locking. The signal and the reference are injected into the interferometer and co-propagate through 8~m of fibers and 0.5~m of free space in each arm. The interferometer is assembled to maximize interference visibility for a broad spectrum and simultaneously provide fast phase switching. At the interferometer outputs, the signal and the reference are separated and detected. The detected reference is processed by a PID controller, and consequently, the phase is locked via a fiber stretcher. Legend: light-emitting diode (LED), interference filter (IF), polarizing beam splitter (PBS), fiber coupler (FC), phase dispersion compensator (PDC), fiber stretcher (FS), electro-optic phase modulator (EOM), quarter-wave plate (QWP), photodiode (PD), sensitive photodiode or single-photon detector (SPD), proportional-integral-derivative controller (PID), analog-to-digital convertor (ADC).}
\label{fig:setup}
\end{figure}

The experimental setup is depicted in Fig.~\ref{fig:setup}. Single-mode coupled light-emitting diode (LED) provides an intensity and spectrally stable source of the reference. The spectral stability is further improved by adding interference filters with a central wavelength of 840~nm and full width at half maximum (FWHM) of 21~nm. In the signal ports, arbitrary light within the 3~nm FWHM around 810~nm can be injected. For active phase locking characterization, we injected the signal in a single input port. The signal is generated by a single-mode coupled and spectrally filtered LED, which perfectly simulates spectral properties of typical quantum photonic sources such as a spontaneous parametric down-conversion pumped by ultra-short pulses. Before coupling into the MZI, both the signal and the reference are horizontally polarized.

\begin{figure}[h!]
\centering\includegraphics[width=0.75\linewidth]{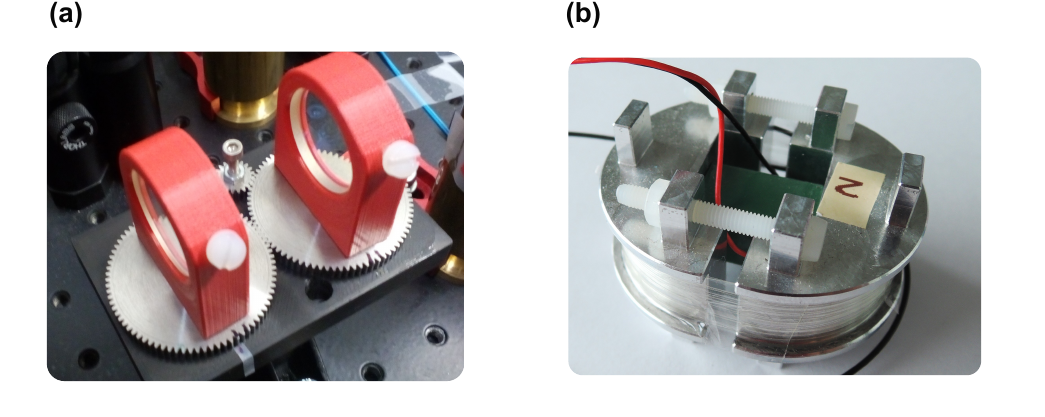}
\caption{(a) Phase dispersion compensator PDC1 used for tuning the relative phase between the signal and the reference. (b) Fiber stretcher used for the phase locking. By applying voltage to a piezoelectric crystal, the fiber coiled around two aluminium segments is stretched and the phase is changed.}
\label{fig:opt_elementy}
\end{figure}

The signal and the reference are merged at a balanced fiber coupler (FC1) and co-propagate through the MZI. To achieve simultaneously high visibility and fast switching capabilities, the MZI combines fiber and free-space elements. Single-mode polarization-maintaining fiber part of a total length of 8~m provides inherent spatial and polarization overlap together with the phase control. Path balancing, polarization filtration, and dispersion compensation are achieved in 0.5~m of free space. Balanced fiber couplers optimized for 810~nm are used to split and merge the interfering light. Their splitting ratios are estimated to be 51:49 for the signal and 35:65 for the reference. The splitting ratio of FC1 can be effectively modified to perfect 50:50 operation by adjusting losses in one MZI arm. The arm lengths are balanced in the free space by micrometric translation of one coupling lens. Thanks to the 21~nm broad spectrum of the reference, we can balance MZI arms within approximately two wavelengths, i.e., with phase precision of $\pm 2\pi$. To improve polarization degree of light, two pairs of polarizing beam splitters (PBSs) with extinction greater than 30~dB are used. One pair of PBSs is placed in the free-space part, and the other pair is included in the wavelength separation part. The MZI allows for fast phase switching via 10~GHz integrated electro-optic modulators (EOMs). This capability is crucial in applications of the MZI, however, it significantly increases chromatic dispersion in the MZI. To cancel out the dispersion, we compose MZI arms symmetrically. Namely, the fiber lengths are balanced with mm precision, and similar EOMs are placed in both arms. Additional dispersion compensation and manipulation are performed in free space with a custom component called phase dispersion compensator consisting of two elements, PDC1 and PDC2. As shown in Fig.~\ref{fig:opt_elementy}(a), the PDC1 consists of two 1~cm thick high-dispersion SF10 glass plates mounted on contra-directional tilting stages. The PDC1 allows for manipulating the relative optical path between the wavelengths without any misalignment of their fiber coupling. Thus we can set an arbitrary relative phase between the signal and the reference. This enables the reference to be locked at the most sensitive point of the interference fringe (phase $\frac{\pi}{2}$) regardless of the signal phase. PDC2 consists of a fixed SF10 glass plate with 3~cm thickness to compensate for the residual dispersion. As a result, we reach the visibility of 99.6$\%$ for both MZI outputs for the 3~nm broad signal and 98.6$\%$ for one output of the 21~nm broad reference.

At each MZI output, the wavelengths are demultiplexed via a dichroic optical isolator assembled from a PBS, a quarter-wave plate (QWP), and an interference filter. We reach negligible crosstalk of 30 photons/s from the reference to the signal, equivalent to extinction better than 80~dB. The separation occurs at the narrow-band interference filter centered at 810~nm, where the signal passes through while the reference is reflected back. Subsequently, the reference is reflected at the PBS. This is possible since the reference polarization is transformed from horizontal to vertical by bi-directional passing through a 45-degree oriented QWP and $\pi$ phase shift caused by reflection at the interference filter. Since the QWP undesirably changes the signal polarization, we use a compensatory QWP after the wavelength separation part. Before the detection, both the signal and the reference are coupled into a single-mode fiber. For the signal, polarization-maintaining fibers are used, which facilitates its transfer to following experimental stages. During active phase locking tests, the signal with the power of about 1~pW ($4\cdot10^6$ photons/s) impinges directly the ultra-sensitive silicon photodiodes with noise equivalent power (NEP) of 1.4~fW/$\mathrm{\sqrt{Hz}}$ and 30~Hz bandwidth. For single-photon measurements, we use single-photon avalanche diodes with a 50 Hz dark count rate plugged into a time tagger. 

The active phase lock is carried out by a feedback loop composed of the photodiodes, PID controller, and a fiber stretcher. The reference with the power of 1.5~nW impinges silicon photodiodes with NEP=9~fW/$\mathrm{\sqrt{Hz}}$ and 2~kHz bandwidth. The resulting electronic signals are processed by an in-house developed analog PID controller based on low-noise operational amplifiers. The circuit is provided with the adaptive setpoint described in Methods. Therefore, we lock the phase regardless of the intensity fluctuations of the reference. The PID controller is fine-tunable in each parameter via 10-turn precision potentiometers. As a result, the target phase is set with high accuracy. To provide correct phase estimation for the adaptive setpoint, we include additional tunability to cancel out detector offsets and to balance detection efficiencies. The PID controller is designed to drive a custom-made fiber stretcher depicted in Fig.~\ref{fig:opt_elementy}(b). It has 7.2~$\mu$F capacitance, 1~kHz bandwidth, 0.11~V half-wave voltage, and dynamic range of $\pm275$~$\mathrm{\mu m}$ \cite{Novakova2020}. Since the extent of the phase drift is much smaller in our case, we use only $\pm110$~$\mathrm{\mu m}$ corresponding to the output voltage range of the PID controller.

To reduce air flux and temperature instabilities, we cover the whole setup with styrofoam plates during all measurements. The passive protection reduces the temperature drift from $2^\circ$C peak-to-peak per 2 hours to a peak-to-peak drift of $0.3^\circ$C per 15 hours.  
Our setup is placed on a 60x60x6~cm breadboard with no additional damping of mechanical vibrations. For the data acquisition, we use an 8-channel 18-bit analog-to-digital converter set to a sampling frequency of 16~Hz. We increase SNR by averaging the data samples to 1~Hz except for the detailed spectral characterization (Fig.~\ref{fig:stabilita}(b)). We collect data at each detector, and additionally, we monitor the voltage applied to the fiber stretcher. The knowledge of the driving voltage, together with the information on half-wave voltage, allows for the evaluation of the compensated phase drift. The collected data are processed by an algorithm based on Eq.~(1). Apart from the output intensities, it requires additional parameters such as the visibility of each MZI output and a ratio of losses in the MZI outputs. For measurements with a non-stabilized phase, these parameters are periodically recovered as interference fringes are scanned spontaneously due to the phase drift. While the stabilization process is running, interference fringe scanning is not possible, thus the additional parameters are independently measured before the stabilization starts.

\section{Results and discussion}
\begin{figure}[h!]
\centering\includegraphics[width=1\linewidth]{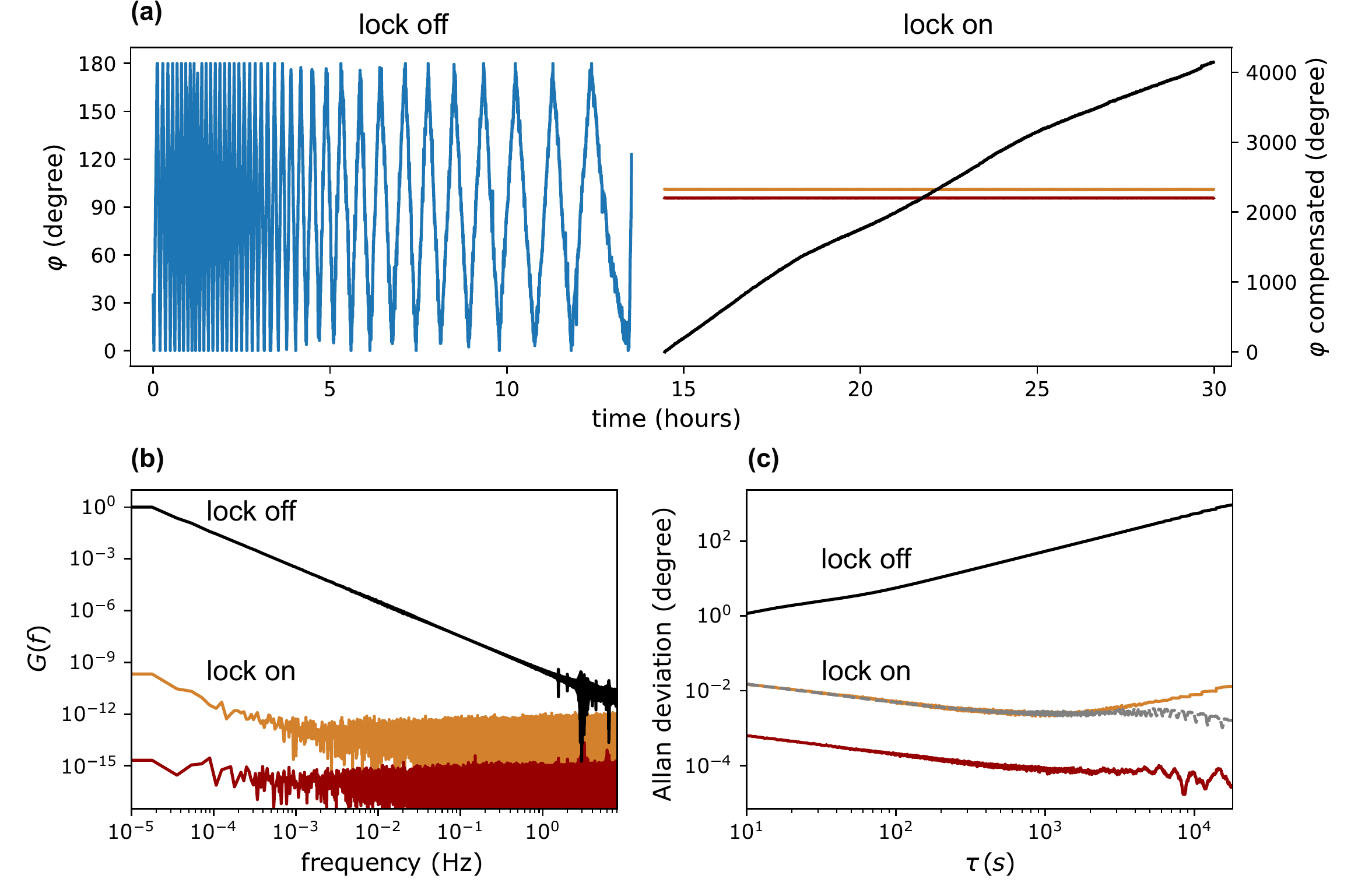}
\caption{Comparison of stabilized and non-stabilized MZI. In stabilized MZI, the reference and the signal are represented by the red line and the orange line, respectively. (a) Stability in the time domain. The blue line represents the phase in non-stabilized MZI and the black line corresponds to the phase drift compensated during the phase locking (see the right vertical axis). (b) Spectral power density $G$ of the phase noise. (c) Allan deviation of the phase noise. The gray line illustrates the noise level of the signal acquisition.}
\label{fig:stabilita}
\end{figure}

We compare the performance of the MZI with and without the active stabilization and present the phase stability in three ways: in the time domain, in the frequency domain, and via the Allan deviation. The results in the time domain are depicted in Fig.~\ref{fig:stabilita}(a). The typical phase drift of non-stabilized MZI is shown on the left side. On the right side, we show phase noise for stabilized MZI (orange and red lines) and the phase compensated during the stabilization process (black line). For further analysis, we use only black, orange, and red datasets since they were achieved within the same environmental conditions. While the phase lock is active, a phase drift of 4000 degrees per 15 hours is virtually eliminated. In terms of standard deviation accumulated over a typical single-photon sampling frequency of 1~Hz, we reach 0.05 degrees for the signal and 0.002 degrees for the reference within a 15-hour interval.
 
Further, we analyze the spectral power density of the phase noise shown in Fig.~\ref{fig:stabilita}(b). During the stabilization process, the phase noise is suppressed by 9 orders of magnitude for frequencies in the range $10^{-3}$ to $10^{-5}$~Hz. For frequencies above $10^{-3}$ Hz, the phase noise is primarily given by detection noise. Furthermore, we evaluate the Allan deviation of the phase noise depicted in Fig.~\ref{fig:stabilita}(c). Allan deviation corresponds to root mean square phase drift for a variable time interval $\tau$ \cite{allan1966}. For the signal, we reach Allan deviation below $2\times10^{-2}$ degrees for time intervals from 10~s to more than 5 hours. The minimum value $2\times10^{-3}$ degrees is achieved for $10^3$~s. Allan deviation of the reference is evaluated to be smaller than $10^{-3}$ degrees. 

The results show that the reference exhibits higher phase precision than the signal. This discrepancy has two reasons. Firstly, the signal is affected by a higher amount of detection noise than the reference. Specifically, the detection noise superimposes the signal up to $10^3$~s as illustrated by the gray line in Fig.~\ref{fig:stabilita}(c). Secondly, the signal is slightly affected by the residual LSD leading to increased phase error in the long term. Considering the aforementioned reasons, the Allan deviation of the signal should be interpreted as an upper bound of the real phase error in the MZI. On the other hand, the Allan deviation of the reference gives the lower bound of the real phase error since it does not reflect LSD. We estimate that in the short term, where LSD is negligible, the real Allan deviation follows the reference curve, whereas, in the long term, it approaches the signal curve. 
The measured phase precision of the reference illustrates the ultimate performance of the phase lock. Therefore, the direct demonstration of sub-$10^{-3}$ degree Allan deviation for time intervals up to $10^4$~s is feasible provided that we completely cancel out LSD and improve SNR of the signal detection.

\begin{figure}[h!]
\centering\includegraphics[width=0.72\linewidth]{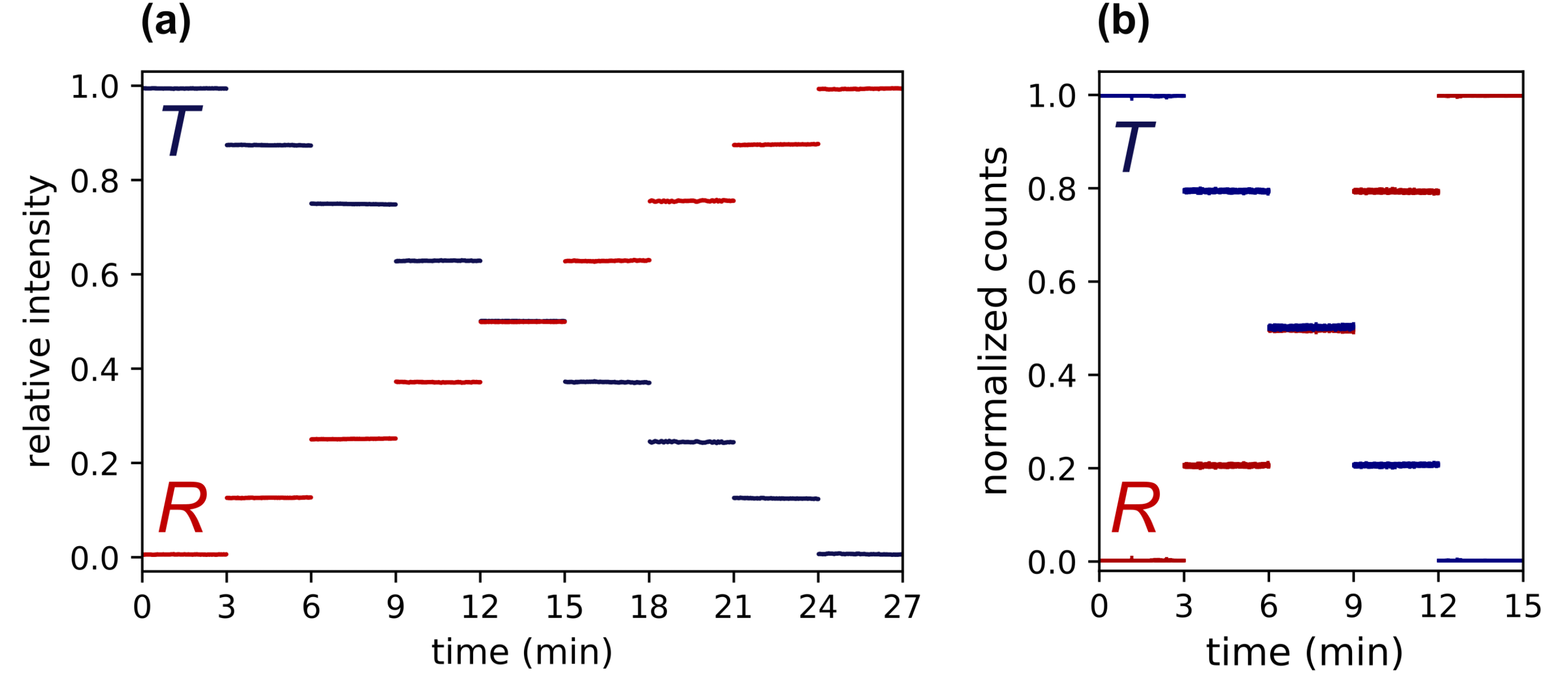}
\caption{Complete phase tunability via the phase dispersion compensator demonstrated for a classical signal (a) and single photons (b). Although the reference is locked at the most sensitive phase $\frac{\pi}{2}$, we are able to tune arbitrary splitting ratio $T$:$R$ for the signal.}
\label{fig:schod}
\end{figure}

Although, in general, the performance of phase locking is dependent on the reference phase (the best performance is achieved in $\pi/2$), we are not limited to a specific phase value of the signal. In Fig.~\ref{fig:schod}, we show the phase locking for an arbitrary signal phase while being locked with the reference close to $\pi/2$. This is possible due to full phase tunability between the signal and the reference achieved by the phase dispersion compensator element (see Experiment for details). Fig.~\ref{fig:schod}(a) shows the full tunability for a classical signal, while in Fig.~\ref{fig:schod}(b) we demonstrate the tunability for the single-photon signal (count rate approx. $5\times10^5$ counts/s, integration time 1~s). Additional results at the single-photon level were achieved previously with preliminary versions of the setup. Firstly, we demonstrated balanced time-bin multiplexing for photon-number resolving detection using a loop configuration \cite{Svarc2019}. Secondly, we demonstrated the Fock state conversion exploiting feedforward control of the interferometer \cite{Svarc2020}. These experiments were performed before a significant improvement of the phase lock, hence they do not reflect the unprecedented phase stability presented here.

Let us compare our results with other state-of-the-art experiments across various interferometer configurations and target applications. Niwa et al. reported 20~pm phase noise between 0.3~mHz and 1~Hz in 5x5 cm free-space MZI \cite{niwa2009}. The results are achieved for strong light and classical detectors. They use advanced passive and active methods of protection against environment changes: the setup is placed in a vacuum chamber with 1~mK thermal stability, and the MZI elements are composed of ultralow expansion glass. In comparison, we reach phase noise around 100~pm within a much larger frequency interval (0.02 mHz to 1 Hz) for an 8~m long fiber interferometer placed in a standard lab environment. In the field of single-photon experiments, the phase stability is typically in order of a few degrees \cite{Makarov2004,Huntington2005,Suzuki2006,Takeno2007,Cho2009,
Xavier2011,Micuda2014,yanikgonul2019}. Five of the aforementioned works are limited to short-term operation (up to few minutes)\cite{Huntington2005,Suzuki2006,Takeno2007,Cho2009,Xavier2011}, whereas long-term stability (up to several hours) is achieved only in three works \cite{Makarov2004,Micuda2014,yanikgonul2019}. To the best of our knowledge, the most precise stabilization of a single-photon interferometer has been reported by Toliver et al. \cite{Toliver2015} and Roztocki et al. \cite{Roztocki2021}. Toliver et al. achieved a 200~s phase lock with a standard deviation of $0.2$ degrees. In contrast, we demonstrate a standard deviation of 0.05 degrees for more than 15 hours. Roztocki et al. reached Allan deviation below 0.2 degrees from $10^{-1}$~s to $10^4$~s, but these results were demonstrated only for a bright reference light (the phase was not reconstructed using an independent signal as in our case). In contrast, we reach Allan deviation below $5\times10^{-3}$ degrees for the reference within the same interval.
 
Our results demonstrate the unprecedented performance of dual-wavelength phase locking across single-photon interferometers. Our approach will work for all types of interferometers and complex interferometric networks, including fiber and free-space propagation. Regarding free-space interferometers, phase locking might be even simpler, as a small spatial displacement between the signal and the reference can substitute the wavelength separation. However, this scheme is not free from the relative phase drift between the signal and the reference since the index of refraction in air locally changes for each of the displaced paths \cite{Micuda2014}. Further, our phase-locking method can be utilized for unbalanced interferometers, provided that  a sufficiently coherent source is employed as the reference. However, highly coherent sources can bring additional challenges due to random interference on optical surfaces and elements, leading to intensity fluctuation throughout the setup.

\section{Conclusion}
We have implemented a novel technique of continuous phase locking for single-photon circuits and networks. Our approach extends dual-wavelength stabilization methods by the adaptive setpoint, enabling precise phase locking despite the fluctuating intensity. We show that the adaptive setpoint, together with thorough optimization in all possible degrees of freedom, leads to unprecedented stability even with faint optical signals. We have experimentally demonstrated sub-0.1 degree phase locking of 8.5~m long fiber-based MZI for more than 15 hours. Specifically, we have reached a phase standard deviation of 0.05 degrees, which is equivalent to the path displacement of about 100~pm. To the best of our knowledge, our results represent the most precise phase locking of single-photon interferometers. Moreover, we have reached complete phase tunability of the device and negligible crosstalk from the reference to the single-photon signal. Since our method can be implemented in a vast majority of interferometric networks, it can significantly improve the performance of many phase-sensitive applications in information processing, quantum metrology, and quantum communication.

\section*{Funding}
Czech Science Foundation (project 19-19189S); Ministry of Education, Youth, and Sports of the Czech Republic (projects CZ.02.2.69/0.0/0.0/19$\_$073/0016713, DSGC-2021-0118).

\section*{Acknowledgments}
We thank Petr Novák for his help with the 3D printing of the phase dispersion compensator mount. Student grant DSGC-2021-0118 is funded under the OPIE project „Improvement of Doctoral Student Grant Competition Schemes and their Pilot Implementation“, 
reg. no. CZ.02.2.69/0.0/0.0/19$\_$073/0016713 in the grant results.

\section*{Disclosures}
The authors declare that there are no conflicts of interest related to this article.

\section*{Data Availability}
Data underlying the results presented in this paper are available in Ref. \cite{SvarcGitHub}.



\begin{thebibliography}{10}
\newcommand{\enquote}[1]{``#1''}

\bibitem{ma2011}
X.~song Ma, S.~Zotter, N.~Tetik, A.~Qarry, T.~Jennewein, and A.~Zeilinger,
  \enquote{A high-speed tunable beam splitter for feed-forward photonic quantum
  information processing,} {\protect\JournalTitle{Opt. Express}} \textbf{19},
  22723--22730 (2011).

\bibitem{mikova2012}
M.~Mikov{\'{a}}, H.~Fikerov{\'{a}}, I.~Straka, M.~Mi{\v{c}}uda,
  J.~Fiur{\'{a}}{\v{s}}ek, M.~Je{\v{z}}ek, and M.~Du{\v{s}}ek,
  \enquote{Increasing efficiency of a linear-optical quantum gate using
  electronic feed-forward,} {\protect\JournalTitle{Phys. Rev. A}} \textbf{85},
  012305 (2012).

\bibitem{Giovannetti2011}
V.~Giovannetti, S.~Lloyd, and L.~Maccone, \enquote{Advances in quantum
  metrology,} {\protect\JournalTitle{Nature Photonics}} \textbf{5}, 222--229
  (2011).

\bibitem{LIGO2013squeezing}
J.~Aasi, J.~Abadie, B.~Abbott, R.~Abbott, T.~Abbott, M.~Abernathy, C.~Adams,
  T.~Adams, P.~Addesso, R.~Adhikari \emph{et~al.}, \enquote{Enhanced
  sensitivity of the {LIGO} gravitational wave detector by using squeezed
  states of light,} {\protect\JournalTitle{Nature Photonics}} \textbf{7},
  613--619 (2013).

\bibitem{Thekkadath2020}
G.~Thekkadath, M.~Mycroft, B.~Bell, C.~Wade, A.~Eckstein, D.~Phillips,
  R.~Patel, A.~Buraczewski, A.~Lita, T.~Gerrits \emph{et~al.},
  \enquote{Quantum-enhanced interferometry with large heralded photon-number
  states,} {\protect\JournalTitle{NPJ Quantum Information}} \textbf{6}, 1--6
  (2020).

\bibitem{Xavier2011}
G.~B. Xavier and J.~P. von~der Weid, \enquote{Stable single-photon interference
  in a 1 km fiber-optic {M}ach--{Z}ehnder interferometer with continuous phase
  adjustment,} {\protect\JournalTitle{Opt. Lett.}} \textbf{36}, 1764--1766
  (2011).

\bibitem{Xavier2015}
G.~Carvacho, J.~Cari{\~{n}}e, G.~Saavedra, {\'{A}}.~Cuevas, J.~Fuenzalida,
  F.~Toledo, M.~Figueroa, A.~Cabello, J.-{\AA}. Larsson, P.~Mataloni, G.~Lima,
  and G.~B. Xavier, \enquote{Postselection-loophole-free {B}ell test over an
  installed optical fiber network,} {\protect\JournalTitle{Phys. Rev. Lett.}}
  \textbf{115}, 030503 (2015).

\bibitem{Toliver2015}
P.~Toliver, J.~M. Dailey, A.~Agarwal, and N.~A. Peters, \enquote{Continuously
  active interferometer stabilization and control for time-bin entanglement
  distribution,} {\protect\JournalTitle{Opt. Express}} \textbf{23}, 4135--4143
  (2015).

\bibitem{Cho2016}
S.-B. Cho and H.~Kim, \enquote{Active stabilization of a fiber-optic two-photon
  interferometer using continuous optical length control,}
  {\protect\JournalTitle{Opt. Express}} \textbf{24}, 10980--10986 (2016).

\bibitem{Xu2019}
Y.~{Xu}, J.~{Lin}, Y.~{Li}, H.~{Dai}, S.~{Liao}, and C.~{Peng}, \enquote{Active
  {P}hase stabilization for the interferometer with 128 actively selectable
  paths,} {\protect\JournalTitle{IEEE Trans. Nucl. Sci.}} \textbf{66},
  1076--1080 (2019).

\bibitem{ma2012}
X.-s. Ma, S.~Zotter, J.~Kofler, R.~Ursin, T.~Jennewein, {\v{C}}.~Brukner, and
  A.~Zeilinger, \enquote{Experimental delayed-choice entanglement swapping,}
  {\protect\JournalTitle{Nature Physics}} \textbf{8}, 479--484 (2012).

\bibitem{Vallone2018}
F.~Vedovato, C.~Agnesi, M.~Tomasin, M.~Avesani, J.-A. Larsson, G.~Vallone, and
  P.~Villoresi, \enquote{Post-selection-loophole-free {B}ell violation with
  genuine time-bin entanglement,} {\protect\JournalTitle{Physical Review
  Letters}} \textbf{121}, 190401 (2018).

\bibitem{Svarc2020}
V.~\v{S}varc, J.~Hlou\v{s}ek, M.~Nov\'{a}kov\'{a}, J.~Fiur\'{a}\v{s}ek, and
  M.~Je\v{z}ek, \enquote{Feedforward-enhanced {F}ock state conversion with
  linear optics,} {\protect\JournalTitle{Opt. Express}} \textbf{28},
  11634--11644 (2020).

\bibitem{Svarc2019}
V.~\v{S}varc, M.~Nov\'{a}kov\'{a}, G.~Mazin, and M.~Je\v{z}ek, \enquote{Fully
  tunable and switchable coupler for photonic routing in quantum detection and
  modulation,} {\protect\JournalTitle{Opt. Lett.}} \textbf{44}, 5844--5847
  (2019).

\bibitem{Takeno2007}
Y.~Takeno, M.~Yukawa, H.~Yonezawa, and A.~Furusawa, \enquote{Observation of -9
  d{B} quadrature squeezing with improvement of phase stability in homodyne
  measurement,} {\protect\JournalTitle{Opt. Express}} \textbf{15}, 4321--4327
  (2007).

\bibitem{Eberle2013}
T.~Eberle, V.~H\"{a}ndchen, and R.~Schnabel, \enquote{Stable control of 10 d{B}
  two-mode squeezed vacuum states of light,} {\protect\JournalTitle{Opt.
  Express}} \textbf{21}, 11546--11553 (2013).

\bibitem{Makarov2004}
V.~Makarov, A.~Brylevski, and D.~R. Hjelme, \enquote{Real-time phase tracking
  in single-photon interferometers,} {\protect\JournalTitle{Appl. Opt.}}
  \textbf{43}, 4385--4392 (2004).

\bibitem{Furusawa2017}
S.~Takeda and A.~Furusawa, \enquote{Universal quantum computing with
  measurement-induced continuous-variable gate sequence in a loop-based
  architecture,} {\protect\JournalTitle{Physical Review Letters}} \textbf{119},
  120504 (2017).

\bibitem{takeda2019}
S.~Takeda, K.~Takase, and A.~Furusawa, \enquote{On-demand photonic entanglement
  synthesizer,} {\protect\JournalTitle{Sci. Adv.}} \textbf{5}, eaaw4530 (2019).

\bibitem{Huntington2005}
D.~Pulford, C.~Robillard, and E.~Huntington, \enquote{Single photon locking of
  an all-fiber interferometer,} {\protect\JournalTitle{Rev. Sci. Instrum.}}
  \textbf{76}, 063114 (2005).

\bibitem{Suzuki2006}
S.~Suzuki, H.~Yonezawa, F.~Kannari, M.~Sasaki, and A.~Furusawa, \enquote{7 d{B}
  quadrature squeezing at 860 nm with periodically poled {K} {T}i {OPO}$_4$,}
  {\protect\JournalTitle{Appl. Phys. Lett.}} \textbf{89}, 061116 (2006).

\bibitem{Cho2009}
S.-B. Cho and T.-G. Noh, \enquote{Stabilization of a long-armed fiber-optic
  single-photon interferometer,} {\protect\JournalTitle{Opt. Express}}
  \textbf{17}, 19027--19032 (2009).

\bibitem{yanikgonul2019}
S.~Yanikgonul, R.~Guo, A.~Xomalis, A.~N. Vetlugin, G.~Adamo, C.~Soci, and N.~I.
  Zheludev, \enquote{Phase stabilization of a coherent fiber network by
  single-photon counting,} {\protect\JournalTitle{Opt. Lett.}} \textbf{45},
  2740--2743 (2020).

\bibitem{Rohde2014}
K.~R. Motes, A.~Gilchrist, J.~P. Dowling, and P.~P. Rohde, \enquote{Scalable
  boson sampling with time-bin encoding using a loop-based architecture,}
  {\protect\JournalTitle{Phys. Rev. Lett.}} \textbf{113}, 120501 (2014).

\bibitem{Pan2017}
Y.~{He et al.}, \enquote{Time-bin-encoded boson sampling with a single-photon
  device,} {\protect\JournalTitle{Phys. Rev. Lett.}} \textbf{118}, 190501
  (2017).

\bibitem{Novakova2020}
M.~Nov{\'{a}}kov{\'{a}}, L.~Podhora, V.~{\v{S}}varc, and M.~Je{\v{z}}ek,
  \enquote{Polarization-maintaining 3{D} printed fiber stretcher,}
  {\protect\JournalTitle{{in preparation}}}  (2022).

\bibitem{allan1966}
D.~W. Allan, \enquote{Statistics of atomic frequency standards,}
  {\protect\JournalTitle{Proceedings of the IEEE}} \textbf{54}, 221--230
  (1966).

\bibitem{niwa2009}
Y.~Niwa, K.~Arai, A.~Ueda, M.~Sakagami, N.~Gouda, Y.~Kobayashi, Y.~Yamada, and
  T.~Yano, \enquote{Long-term stabilization of a heterodyne metrology
  interferometer down to a noise level of 20 pm over an hour,}
  {\protect\JournalTitle{Appl. Opt.}} \textbf{48}, 6105--6110 (2009).

\bibitem{Micuda2014}
M.~Mi{\v{c}}uda, E.~Dol{\'{a}}kov{\'{a}}, I.~Straka, M.~Mikov{\'{a}},
  M.~Du{\v{s}}ek, J.~Fiur{\'{a}}{\v{s}}ek, and M.~Je{\v{z}}ek, \enquote{Highly
  stable polarization independent {M}ach-{Z}ehnder interferometer,}
  {\protect\JournalTitle{Rev. Sci. Instrum.}} \textbf{85}, 083103 (2014).

\bibitem{Roztocki2021}
P.~Roztocki, B.~MacLellan, M.~Islam, C.~Reimer, B.~Fischer, S.~Sciara,
  R.~Helsten, Y.~Jestin, A.~Cino, S.~T. Chu \emph{et~al.}, \enquote{Arbitrary
  phase access for stable fiber interferometers,} {\protect\JournalTitle{Laser
  \& Photonics Reviews}} \textbf{15}, 2000524 (2021).

\bibitem{SvarcGitHub}
{V.Švarc}, \enquote{{Git{H}ub repository
  {S}varc{O}ptics{O}lomouc/stabilization},}
  \url{https://github.com/SvarcOpticsOlomouc/stabilization}.

\end{thebibliography}
\end{document}